\documentclass{an}
\usepackage{graphicx}
\usepackage{times}
\usepackage{fancyhdr}
\usepackage{eurosym}
\usepackage{psfig}
\def\HI{H\,{\sc i}}
\def\HII{H\,{\sc ii}}

\def\etal{{et~al.}}

\begin{document}

\title{The Square Kilometre Array: A new probe of cosmic magnetism}

\author{Bryan M. Gaensler\inst{1,2}}
\institute{
Harvard-Smithsonian Center for Astrophysics,
60 Garden Street MS-6, Cambridge 02138, USA
\and
Project Scientist, International SKA Project Office}

\date{Received; accepted; published online}

\abstract{Magnetic fields are a fundamental part of many
astrophysical phenomena, but the evolution, structure and origin of
magnetic fields are still unresolved problems in physics and astrophysics.
When and how were the first fields generated? Are present-day magnetic
fields the result of standard dynamo action, or do they represent
rapid or recent field amplification through other processes? What role
do magnetic fields play in turbulence, cosmic ray acceleration and
structure formation?  I explain how the Square Kilometre Array (SKA),
a next-generation radio telescope, can deliver stunning new data-sets
that will address these currently unanswered issues.  The foundation
for these experiments will be an all-sky survey of rotation measures,
in which Faraday rotation toward $>10^7$ background sources will provide
a dense grid for probing magnetism in the Milky Way, nearby galaxies,
and in distant galaxies, clusters and protogalaxies.  Using these data,
we can map out the evolution of magnetized structures from redshifts $z >
3$ to the present, can distinguish between different origins for seed
magnetic fields in galaxies, and can develop a detailed model of the
magnetic field geometry of the intergalactic medium and of the overall
Universe.  In addition,
the SKA will certainly discover new magnetic phenomena beyond what we
can currently predict or imagine.
}

\correspondence{bgaensler@cfa.harvard.edu}

\maketitle

\section{Introduction}

The Square Kilometre Array\footnote{See http://www.skatelescope.org
for more information.} (SKA) is a next generation radio telescope, which
will have a collecting area of $\sim10^6$~m$^2$.  With these capabilities,
the SKA will be able to answer fundamental questions about the origin
and evolution of the Universe.

The international SKA consortium, comprised of representatives from
16 member countries, is currently considering site proposals from
Argentina/Brazil, Australia, China, and South Africa. A shortlist will
be selected from these site proposals in the second half of 2006.

The specifications for the SKA (Jones 2004\nocite{jon04}) require
an angular resolution of $0\farcs02$ at 1.4~GHz, a frequency
capability of 0.1--25~GHz, and a field of view of at least 1~deg$^2$
at 1.4 GHz.  While the detailed design for the SKA is yet to be
finalised, the SKA Reference Design (Schilizzi 2006\nocite{sch06}) 
currently consists of a
central 5-km core of both steerable dishes and passive aperture
tiles, with 50\% of the total collecting area distributed on longer
baselines, extending out to $>3000$~km. The dishes will be outfitted
with phased arrays at low frequencies and wide-band feeds at higher
frequencies. The aperture array will have a very wide field of view
at low frequencies ($\sim50$~deg$^2$ at 700~MHz), to allow rapid
surveys of redshifted \HI. Operations for the SKA are expected to
begin in 2015--2020, with a total cost for the instrument of
approximately \euro 1.5 billion. An illustration of the layout of
the Reference Design is shown in Figure~\ref{fig_ref}.

\begin{figure*}
\begin{minipage}{\textwidth}
\vspace{10cm}
\caption{An artist's impression of the SKA Reference Design, showing
the central core of dishes and tiles, surrounded by five spiral arms
containing stations of dishes. 
Image created by XILOSTUDIOS for the International SKA Project
Office.}
\label{fig_ref}
\end{minipage}
\end{figure*}

Five key science projects have been adopted by the SKA International
Steering Committee:
\begin{itemize}

\item The Cradle of Life (Lazio, Tarter \& Wilner 2004\nocite{ltw04});

\item Strong Field Tests of Gravity Using Pulsars and Black Holes (Kramer
\etal\ 2004\nocite{kbc+04});

\item Probing the Dark Ages (Carilli \etal\ 2004a\nocite{cfb+04});
 
\item Galaxy Evolution, Cosmology and Dark Energy (Rawlings \etal\
2004\nocite{rab+04});

\item The Origin and Evolution of Cosmic Magnetism (Gaensler, Beck \& Feretti
2004\nocite{gbf04}).
\end{itemize}
In addition, the theme of ``Exploration of the Unknown'' (Wilkinson
\etal\ 2004\nocite{wke+04}),
which emphasises radio astronomy's outstanding track record of
serendipitous discovery, has been adopted as an underlying philosophy
for design and costing.

\section{Fundamental physical questions}

One of the criteria for the key science projects listed above is
that they must address important but currently unanswered questions
in fundamental physics or astrophysics. For the ``Cosmic Magnetism''
project, indeed many such questions arise. We can broadly group these
questions into three themes, relating to the structure, evolution and
origin of magnetic fields.

The most direct goal is to understand the structure of celestial magnetism
on all scales, by mapping the geometry and strength of magnetism in the
interstellar medium (ISM), intercluster medium (ICM) and intergalactic
medium (IGM), all of which are currently poorly characterised, or in the
case of the IGM, still yet to be definitively detected.  Coupled to these
measurements, we lack an understanding of the way in which magnetic
fields connect large-scale, ordered flows to small-scale, turbulent
flows. Measurements of magnetic fields on all scales are essential to
addressing this issue.

Once the field structure has been determined, we can then ask how these
magnetic fields evolve.  Specifically, we would like to understand the
details of how magnetic fields are generated and then maintained over
cosmic epochs, in both individual galaxies and in galaxy clusters.
We also need to understand the feedback between galaxy evolution and
the associated magnetic fields.

Finally, there are the difficult but key questions relating to the origin
of magnetic fields. As discussed extensively at this meeting, we have yet
to establish if seed fields for galaxies and clusters were primordial,
or if they were injected at early epochs by stars, supernovae and AGN.
Did magnetic fields trace, or even regulate, structure formation in the
early Universe? And ultimately, how and when were the first magnetic
fields generated? Major advances on all these questions can be achieved
with the SKA.

\section{Faraday rotation with the SKA}

Faraday rotation will be a crucial component of SKA studies of the
magnetic Universe. Not only will the SKA's large collecting area allow the
detection of Faraday rotation toward much fainter sources than is possible
now, but the high signal-to-noise ratio and large continuum bandwidth will
yield accurate rotation measures (RMs) and Faraday-corrected polarisation
position angles from a single observation. This is in contrast to current
observations, where such measurements often require multiple observations
at widely separated frequencies. This approach is not only time-consuming, but
may be misleading due to the strong frequency dependence of internal
Faraday effects. 

For example, in a one hour observation of a 0.1~mJy source with a
fractional linear polarisation of 1\%, the SKA will make a 10-$\sigma$
detection of linear polarisation, resulting in an RM error of just
5~rad~m$^{-2}$ and an error in intrinsic polarisation position angle of
only $10^\circ$. Spectacular images of rotation measure and intrinsic
polarisation vectors in supernova remnants, galaxies and clusters can
result, all contemporaneous with standard continuum observations.

\section{The polarised sky with the SKA}

The polarisation of diffuse emission can be difficult to interpret, due
to complicated resolution- and frequency-dependent Faraday effects (e.g.,
Sokoloff \etal\ 1998\nocite{sbs+98}).
Thus the simplest measurements to make are those toward compact
sources which, provided that intrinsic Faraday rotation is minimal,
provide a simple measure of the foreground RM along the line-of-sight.

Currently approximately 2000 compact extragalactic sources and 550
pulsars have measured RMs (Xu \etal\ 2006\nocite{xkhd06}; Han \etal\
2006\nocite{hml+06}, and references therein). Over the last several
decades, these data have proved to be very powerful probes of magnetic
fields in sources ranging from distant Ly-$\alpha$ absorbers to the solar
wind. However, as mentioned above, many of these measurements entailed
polarisation measurements at widely separated frequencies, resulting both
in slow progress and in the possibility of erroneously calculated RMs.
Furthermore, the sampling of these RM data is quite sparse --- over most
of the sky the density of extragalactic RMs is $\sim0.05$~deg$^{-2}$.
The situation is somewhat better in the Galactic plane, where recent
surveys have identified many hundreds of extragalactic and pulsar RMs
(Brown, Taylor \& Jackel 2003\nocite{btj03}; Han \etal\ 2006\nocite{hml+06}),
bringing the density of measurements at low latitudes up to about
$\sim2$~deg$^{-2}$.

To predict the improvements on these data that the SKA can provide, we
need to extrapolate. Beck \& Gaensler (2004\nocite{bg04}) 
have convolved an estimate of the probability
distribution of fractional polarisation for extragalactic sources with
standard $\log N - \log S$ differential source count models to derive a
distribution of ``$\log N - \log P$'', predicting the density of polarised
sources and RM measurements on the sky.

\begin{figure*}
\begin{minipage}{\textwidth}
\centerline{\psfig{file=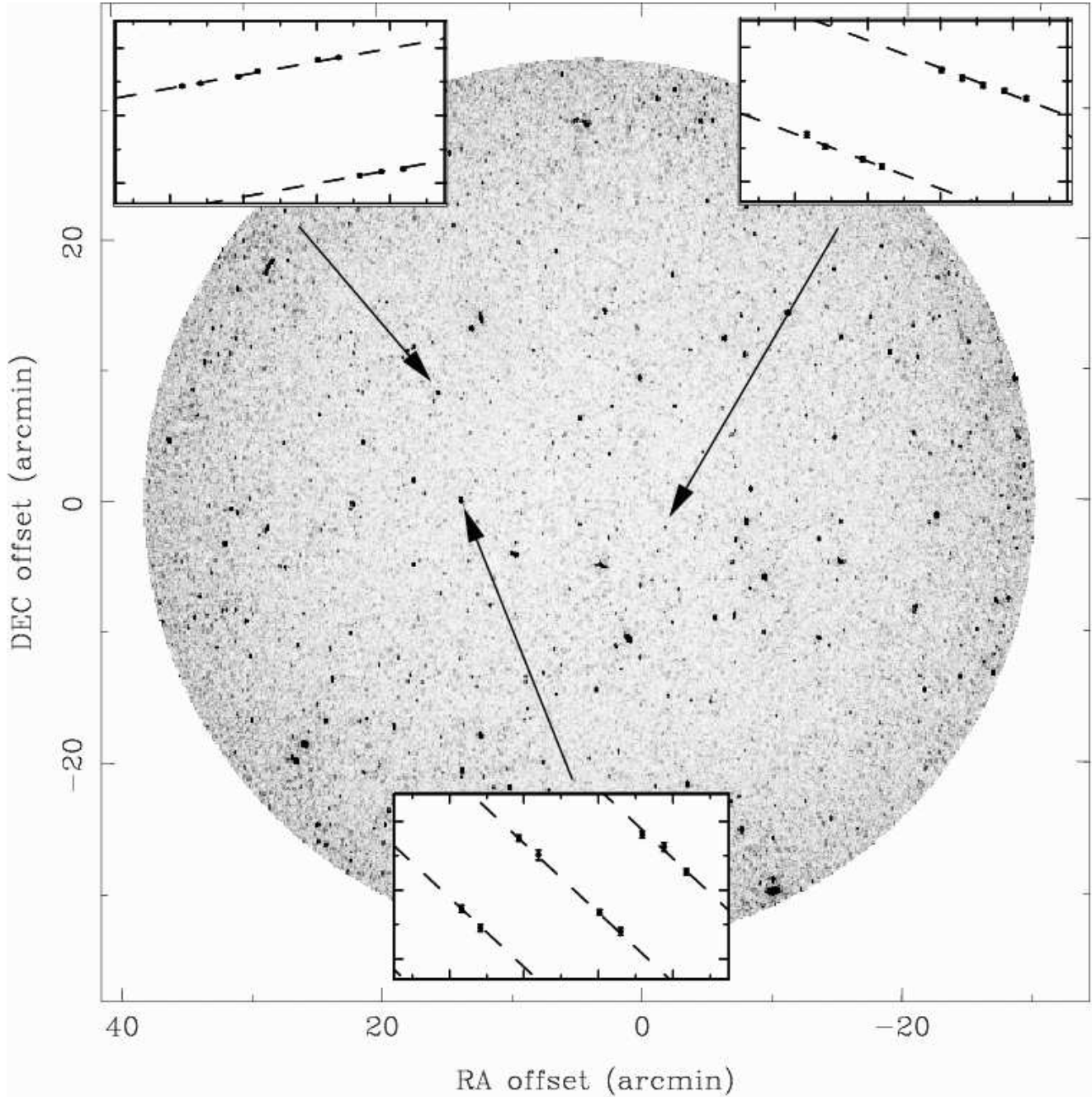,width=\textwidth}}
\caption{A depiction of how the polarised sky at 1.4~GHz will appear
to the SKA after a 5-min integration, based on the
differential source count predictions of Beck
\& Gaensler (2004\protect\nocite{bg04}). The greyscale shows linearly
polarised intensity over a 1-deg$^2$ field-of-view, with a correction
for primary beam attenuation applied.  The panels show polarisation
position angle vs.\ wavelength squared for three compact sources in
the field. A linear fit to each data-set is shown, the slope of which
yields the RM. (This image is adapted from total intensity images of
the Phoenix Deep Survey; Hopkins \etal\ 2003\protect\nocite{hac+03}.)}
\label{fig_sim}
\end{minipage}
\end{figure*}

An example of what these calculations predict is shown in
Figure~\ref{fig_sim}, where we illustrate what a five minute
integration with the SKA might produce. In the 1~deg$^2$ field shown
here, RMs can be detected for polarised fluxes as faint as 3~$\mu$Jy,
resulting in $\approx$500 RMs~deg$^{-2}$, with an average separation
between measurements of $2'-3'$.  Clearly any magnetised extended
source in the foreground, whether directly emitting in the radio
band or not, will be detected and probed by this ``RM grid'',
allowing entirely new ways of studying magnetic fields in distant
objects (see Gaensler \etal\ 2005\nocite{ghs+05}).

We correspondingly envisage a wide-field SKA survey for RMs, in which
$10\,000-20\,000$~deg$^2$ would be imaged in 1.4~GHz continuum emission
and polarisation down to an RMS sensitivity of 0.1~$\mu$Jy.  For a
1-deg$^2$ field-of-view, this survey would require about 12 months of
telescope time. The total intensity component of such a survey
would also be an incredible resource for many other science projects,
as demonstrated by the highly successful FIRST survey (Becker,
White \& Helfand 1995\nocite{bwh95})
(which unfortunately does not include polarisation).

Applying the $\log N - \log P$ function described above, we predict that
this would yield $\sim(2-5)\times10^7$ RMs over the sky, with a mean
spacing between measurements of $\sim90''$.  In addition, virtually all
the detectable radio pulsars in the Galaxy 
would be seen through
their time-averaged emission, yielding an additional $\sim20\,000$ pulsar
RMs, concentrated in the Galactic plane
(Cordes \etal\ 2004\nocite{ckl+04}). 
In the following sections,
we describe some of the scientific yields of the RM grid and
of other polarisation measurements with the SKA
(see Gaensler \etal\ 2004; Beck \& Gaensler 2004; Feretti, Burigana
\& En{\ss}lin 2004;
Feretti \& Johnston-Hollitt 2004,\nocite{gbf04,bg04,fba04,fj04}
for further details and other possible projects).

\section{The magnetic fields of galaxies and clusters}

\subsection{The Milky Way}

Optical and radio polarisation studies have established that the
Milky Way and many other nearby spiral galaxies all show well-organised,
large-scale magnetic fields (Beck 2005\nocite{bec05}).  These
coherent magnetic fields can be generated and preserved through the
dynamo mechanism, in which small-scale turbulent fields are steadily
amplified and ordered by differential rotation (Ruzmaikin, Sokolov \&
Shukurov 1988\nocite{rss88}; Beck \etal\ 1996\nocite{bbm+96}).  However, dynamos
are not yet well understood and still face theoretical difficulties
(Kulsrud 1999\nocite{kul99}).

Our own Milky Way is an excellent test-bed to address these issues,
its large extent on the sky providing a huge ensemble of RMs which can
be used to probe its three-dimensional magnetic field structure. Indeed,
RMs for pulsars and for extragalactic sources have yielded the strength
and orientation of the local regular magnetic field in the plane ($B\sim2$~$\mu$G,
directed approximately azimuthally), have allowed us to identify the
overall geometry of the spiral magnetic pattern of our Galaxy, and have
revealed the surprising presence of large-scale ``magnetic reversals''
(see Crutcher, Heiles \& Troland 2003\nocite{cht03};
Shukurov 2005\nocite{shu05}; Beck 2006\nocite{bec06} for reviews).  
Unfortunately, the sparse sampling of
RMs, and the paucity of pulsar polarisation measurements at distances
larger than $\sim 6-8$~kpc, make it difficult to establish any consensus.

The SKA provides the opportunity to dramatically improve this
situation.  As mentioned above, RMs will be obtained for $\sim20\,000$
pulsars, primarily concentrated in the Galactic plane.
For many of these pulsars, good distance estimates
will be obtained through either either astrometric parallaxes or \HI\
absorption. The resulting dispersion measures and RMs can provide a
comprehensive three-dimensional model of magnetic fields and ionised gas
in the disk and spiral arms (e.g., Stepanov \etal\ 2002\nocite{sfss02}).
On smaller scales, these data can also provide detailed measures of
magnetic fields in individual sources such as supernova remnants and
\HII\ regions (Gaensler \etal\ 2001\nocite{gdm+01}; Beck \& Gaensler
2004\nocite{bg04}), and can characterise the role of magnetic fields
in interstellar turbulence (Lazio, Spangler \& Cordes 
1990\nocite{lsc90}; Minter \& Spangler 1996\nocite{ms96b}).
Alongside these measurements, the SKA will also produce
spectacular images of diffuse polarised emission from individual
sources and from the overall Galactic disk. By considering the
appearance of these structures as a function of frequency,
``Faraday tomography'' can be performed, in which individual
magneto-ionic structures along the line of sight can be
isolated and studied (see Beck \& Gaensler 2004\nocite{bg04}).

Also of considerable interest is the magnetic field structure in the
Galactic halo, and in the  Galactic plane but at
radii beyond the stellar disk
where most pulsars are found.  The dense background grid of extragalactic
RMs can be used to map these regions, providing key information on the
parity and overall geometry of the field.

\subsection{Nearby galaxies and clusters}

\begin{figure}[h!]
\centerline{\psfig{file=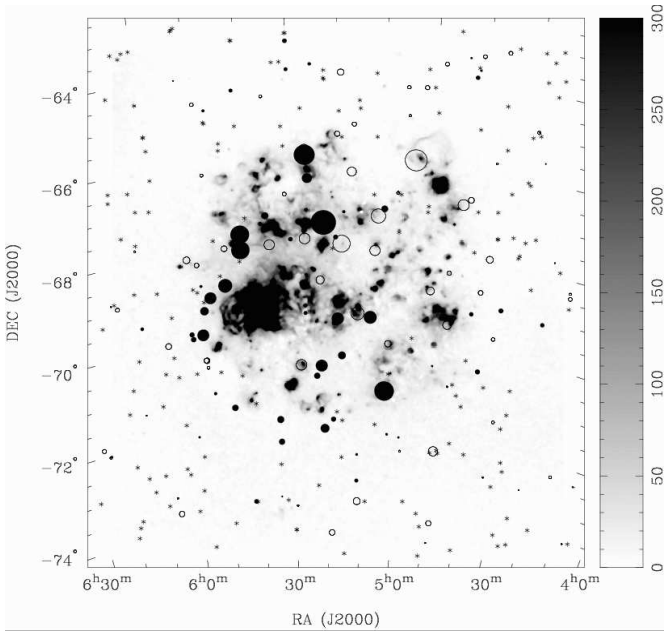,width=0.5\textwidth}}
\caption{Faraday rotation measures through the Large Magellanic Cloud
(Gaensler \etal\ 2005\protect\nocite{ghs+05}).  
The image shows the distribution of extinction
corrected emission measure toward the LMC in units of pc~cm$^{-6}$,
derived from the H$\alpha$ survey of Gaustad \etal\
(2001\protect\nocite{gmrv01}).
The symbols show the sign and magnitude of the RM at various
positions (after subtraction of a mean baseline).  Filled and open
circles correspond to positive and negative RMs, respectively, while
asterisks indicate RMs which are consistent with zero within their errors.
The diameter of each circle is proportional to the magnitude of the RM,
the largest positive and negative RMs being $+247\pm13$~rad~m$^{-2}$
and $-215\pm32$~rad~m$^{-2}$, respectively.}
\label{fig_lmc}
\end{figure}

The same techniques as used for our Milky Way can be applied to more
distant galaxies and clusters. Until recently, such efforts have been
severely limited: Han, Beck \& Berkhuijsen (1998\nocite{hbb98}) 
were able to find just 21 sources with RMs behind
M~31, while Govoni \etal\ (2001\nocite{gtd+01}) 
were able to identify only six
sources with RMs for the
cluster Abell~514. What the SKA will be able to provide is suggested by the
recent RM survey of the Large Magellanic Cloud (LMC) by Gaensler
\etal\ (2005\nocite{ghs+05}), in which
about 100 background RMs were identified, as shown in Figure~\ref{fig_lmc}.  
This dense
sampling allowed the first detailed studies of the LMC's magnetic
field, revealing a coherent axisymmetric spiral field, on which large
fluctuations on all scales are superimposed. This result provides clear
evidence that field amplification in galaxies can be extremely rapid.

With the SKA, this same approach can be extended to many other systems.
For the nearest galaxies, deep SKA observations will provide $>10^5$
background RMs, and thus will yield fantastically detailed maps of the
magnetic structure. Several hundred more galaxies will have $\sim50$ RMs
behind them.  These data will provide maps of galactic magnetic fields
to large galactic radii for a wide range of inclinations, galaxy types
and environments. These statistics will provide stunning new constraints
for dynamo and other theories (see Beck, these proceedings).

In galaxy clusters, magnetic fields regulate heat conduction, and regulate
cluster formation and evolution. There are a variety of ways of measuring
the field strength (Carilli \& Taylor 2002\nocite{ct02}), but measurements
of the field geometry can come only from RMs of background or embedded
sources (for which there are typically $<5$ sources per cluster), and from
measurements of the polarisation position angle of diffuse emission from
the cluster itself (which is usually of very low surface brightness).
With the SKA, the RM grid can provide $\sim1000$ background RMs behind
a typical cluster, while continuum mapping with the core of the array
will detect extended polarisation from both relic and
halo components (e.g., de Bruyn \& Brentjens 2005\nocite{db05}; Govoni \etal\
2005\nocite{gmf+05}).  One can
apply the technique of ``RM synthesis'' to these data, in which a cube
of polarisation vs. wavelength squared is Fourier transformed to yield
the signal as a function of Faraday depth (Brentjens
\& de Bruyn 2005\nocite{bd05}). Through this approach,
a three-dimensional dissection of the cluster's magnetic structure can be
derived.  Furthermore, detailed comparisons between RMs and X-ray emission
in clusters will be possible, allowing us to relate the efficiency of
thermal conduction to the magnetic properties of different regions,
and to directly study the interplay between magnetic fields and hot gas.

For both clusters and individual galaxies, further information can be
obtained when the polarised sources against which RMs are being derived
are extended. Over limited regions, this then provides a map of RM
in the foreground source at the full resolution of the observations,
from which the magnetic field power spectrum in the interstellar and
intercluster medium can be derived (e.g., Vogt \& En{\ss}lin
2003\nocite{ve03}).

\subsection{Polarisation silhouettes}

\begin{figure*}
\begin{minipage}{\textwidth}
\centerline{\psfig{file=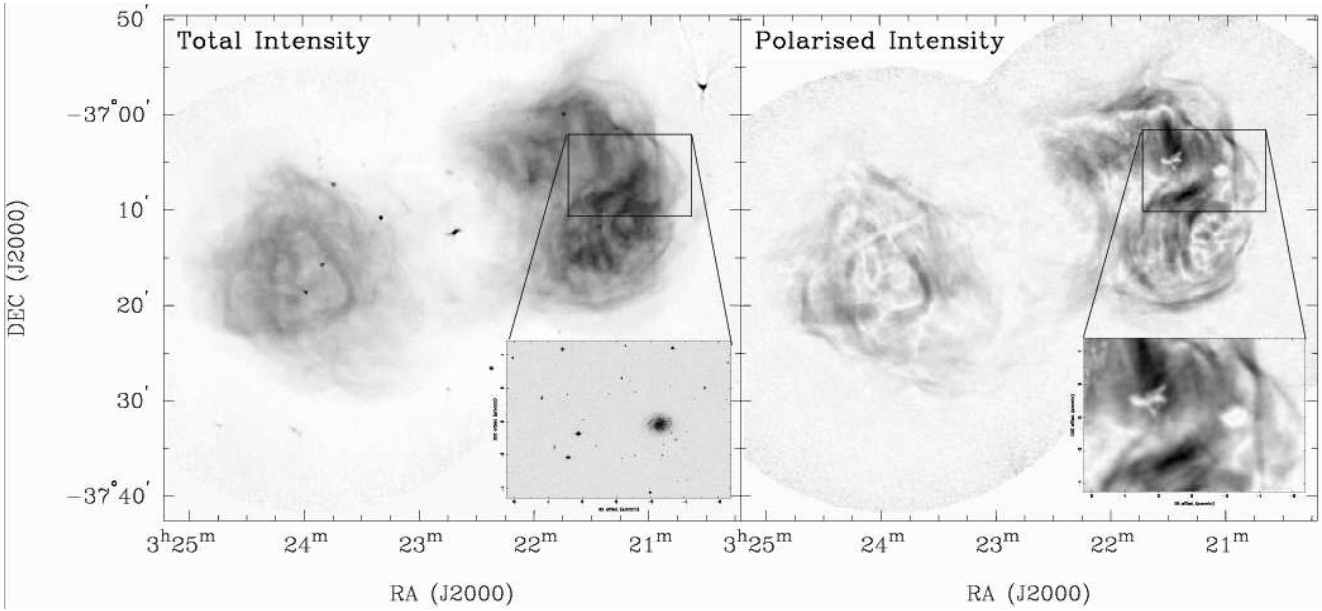,width=\textwidth}}
\caption{Comparison of total intensity (left) and polarised (right)
radio emission from the giant radio galaxy Fornax~A (Fomalont
\etal\ 1989\protect\nocite{feve89}). The insets
show a close-up of the western lobe. The left inset shows
optical emission from the Digitized Sky Survey,
in which NGC~1310 can clearly be seen. The right inset shows linear
polarisation, in which NGC~1310 clearly depolarises the lobe behind it.}
\label{fig_fornax}
\end{minipage}
\end{figure*}

At intermediate redshifts, galaxies and clusters become too small to
be usefully probed by the RM grid. However, one can study magnetic
fields in these sources when they lie in front of distant,
extended, polarised sources. The foreground Faraday rotation then
produces a ``polarisation silhouette''. A good example of this
is NGC~1310, a spiral galaxy
which has depolarised a small part of the radio lobe of Fornax~A, as shown
in Figure~\ref{fig_fornax} (Fomalont \etal\ 1989\nocite{feve89}). 
The RM and fractional polarisation as a function of
both position and frequency toward NGC~1310 can then be used to derive
both the coherent and ordered components of this galaxy's magnetic field
(Schulman \& Fomalont 1992\nocite{sf92}).

While NGC~1310 is reasonably nearby, this same technique can readily be
applied to more distant sources. Several examples of such silhouettes
have already been identified in the literature (e.g., Kronberg,
Perry \& Zukowski 1992\nocite{kpz92}; Johnson, Leahy \& Garrington 1995\nocite{jlg95}). 
With the SKA,
observations of high resolution and sensitivity can identify many such
sources, at a range of redshifts. These data then become a powerful probe
of the evolution of galactic magnetism over the last few billion years.

\subsection{Lyman-$\alpha$ absorbers at $z \sim 1-3$}
\label{sec_lyman}

Most quasars have many foreground clouds along the line-of-sight, as
evidenced by the Ly-$\alpha$ forest seen in their optical spectra.
If both the RM and the redshift of a quasar is known, then the
correlation of RM vs.\ redshift, averaged over a large sample of quasars,
should directly trace the evolution of magnetic fields in galaxies and
proto-galaxies out to large distances (e.g., Welter, Perry \& Kronberg
1984\nocite{wpk84}).  This experiment has been
attempted several times with existing data sets, but the small sample
size, combined with the difficulty of accounting for the foreground RM
contribution from the Milky Way, has meant that these data provide only
marginal (if any) evidence for any evolution of RM with redshift (Perry,
Watson \& Kronberg 1993\nocite{pwk93}; Oren \& Wolfe 1995\nocite{ow95}).

In the future, we expect dramatic improvements. First, the SKA should
be able to identify RMs toward many millions of quasars, a much larger
sample than is available now.  Second, it should be straightforward to
accurately remove the spatially varying foreground contribution to the RM,
because of the very dense sampling of other RM measurements projected near
each quasar. Finally, by combining these data with the wide-field
spectroscopic surveys planned with WFMOS, and with all-sky multi-band
photometry with LSST and SkyMapper, redshifts (and for spectroscopic
studies, information on the number and depth of absorbing systems) can
be obtained for a large fraction of the quasars with RMs. The result
will be a detailed probe of how magnetic fields evolve in galaxies and
their progenitors out to moderate redshifts.

\section{The magnetised intergalactic medium}

It is quite likely that the overall IGM is magnetised, although direct
detection of these fields has so far been difficult, with upper limits
in the range $|B_{\rm IGM}| \la 10^{-8}-10^{-9}$~G 
(Kronberg 1994\nocite{kro94}; Blasi, Burles \& Olinto 1999\nocite{bbo99};
Jedamzik, Katalini\'{c} \& Olinto 2000\nocite{jko00}).
This magnetic field may represent the seed field for
galaxies and clusters, and may play an important role in reionisation
and in the formation
of large-scale structure (Wasserman 1978\nocite{was78}; Sethi
\& Subramanian 2005\nocite{ss05}).  The origin of magnetism in the IGM is
unclear: it may be a primordial field formed in the very early Universe,
it may be injected into the IGM by stars or active galaxies, or it may be
generated in the shocks and turbulence produced in clusters and supernova
remnants (Furlanetto \& Loeb 2001\nocite{fl01}; Kronberg, these proceedings;
Hanayama \etal, these proceedings; Fujita \& Kato, these proceedings).

In principle, the magnetic field of the IGM can be identified through RM
measurements of distant sources. Specifically, if there are large-scale
magnetic fields on a particular scale, then given sufficient statistics,
the angular correlation function of RMs for a given redshift bin should
show a signal (Kolatt 1998\nocite{kol98}). For a range of scales and
a range of redshifts, the magnetic power spectrum of the IGM can then be
determined (Blasi \etal\ 1999\nocite{bbo99}).

There are various difficulties with this experiment, including the need to
accurately remove the Galactic foreground RM (as discussed in
Sect.~\ref{sec_lyman}),
and the small sample size of RMs currently available.  Furthermore, the
signal is easiest observed at higher redshifts, where co-moving magnetic
field strengths and electron densities are presumed to be higher. For an
RM sample dominated by sources with redshifts $z \la 0.5-1$, identifying
this signal is difficult.

We thus envisage observations of very deep polarisation fields with the
SKA, in which a large number of RMs at a variety of redshifts would
be identified (the total intensity component of these data would have
many other applications; e.g., Jackson 2004\nocite{jac04}). 
Accompanying optical and infrared surveys
could provide identifications 
the polarised sources, and could determine their redshifts. With
such a data-set in hand, we can expect to directly detect the IGM field,
and determine its strength, structure, and characteristic length scales.

Several authors have modeled the magnetised
intergalactic shocks which trace the
large-scale structure of the Universe, and which should themselves emit
in synchrotron emission (Keshet, Waxman \& Loeb 2004a\nocite{kwl04a}; 
Br\"{u}ggen
\etal\ 2005\nocite{brs+05}).  Some of this structure has been hinted at
in observations (Bagchi \etal\ 2002\nocite{bem+02};
Kronberg, these proceedings). The SKA will
have the capability to carry out very sensitive, wide field surveys of
emission at low frequencies. If foregrounds can be accurately removed,
we may thus be able to map these shocks in total intensity, providing
a direct estimate of the strength and degree of ordering of magnetic
fields in these structures (Keshet,  Waxman \& Loeb
2004b\nocite{kwl04b}).  If these shocks can also be detected in
linear polarisation, the three-dimensional geometry of the magnetic
field (and presumably of the underlying structure) can be determined.
Brentjens \& de Bruyn (2005\nocite{bd05}) 
have demonstrated the utility of RM synthesis at low frequencies,
where there is good sensitivity to very small changes in RM. Application
of this technique to polarised large-scale structure may be feasible.

\section{Magnetic fields at high redshift}

As discussed by Zweibel (these proceedings) and by Kronberg (these
proceedings), there is good evidence for the existence of microgauss
strength magnetic fields at redshifts $z\sim1-2$. If these measurements
can be extended to redshifts $z>5$, the strength of the field at these
early epochs may provide constraints on how the field was created and
then amplified. 

The SKA and its pathfinders are expected to identify many polarised
radio sources at high redshift, for which RMs can be measured ---
examples include gamma-ray burst afterglows (e.g., GRB~050904 at
$z=6.29$; Kawai \etal\ 2005\nocite{kyk+05}) and distant radio galaxies
(e.g., SDSS~J1148+5251 at $z=6.43$; Carilli \etal\ 2004b\nocite{cwb+04}).
Since the RMs toward these sources represent an integral along the entire
sight-line, foreground components need to be removed to isolate the RM
contribution at high $z$. This can be achieved by deep radio and optical
observations of these fields, which will identify many polarised sources
in close angular proximity to the target source, but at lower redshifts.

At much higher redshifts, RMs may be detectable against the polarised
signal from the cosmic microwave background (e.g., Kosowsky \etal\
2005\nocite{kklr05}). However, this experiment could be challenging,
because of the small position angle changes expected at high frequencies
($\nu > 20$~GHz).

\section{Polarisation pathfinders for the SKA}

While the full SKA is perhaps a decade or more in the
future, a number of polarisation pathfinder experiments are
now being built, which will begin to characterise 
the polarised sky, and which can consequently be used to
explore some of the topics discussed above. These
efforts include:
\begin{itemize}
\item The Galactic Arecibo L-Band Feed Array Continuum Transit Survey
(GALFACTS),\footnote{http://www.ras.ucalgary.ca/GALFACTS} a 1.4-GHz
survey to begin later in 2006 to map the entire
polarised sky visible to Arecibo;
\item The Low Frequency Array (LOFAR),\footnote{http://www.lofar.org} currently under
construction in the Netherlands and Germany, which will
study polarisation over the whole northern sky
at very low frequencies ($\nu =30--80, 110--240$~MHz);
\item The Allen Telescope Array
(ATA),\footnote{http://astron.berkeley.edu/ral/ata} currently being constructed
in northern California, which will have a wide field
of view (5~deg$^2$ at 1.4~GHz) and can carry out very large continuum surveys;
\item The Square Kilometre Array Molonglo Prototype (SKAMP),
a refurbishment of the Molonglo Observatory Synthesis Telescope in south-eastern
Australia, which will provide 18\,000~m$^2$ of collecting area for
studying diffuse polarisation at frequencies $\sim1$~GHz over wide fields;
\item The Low Frequency Demonstrator component of the
 Mileura Wide Field Array
(MWA),\footnote{http://web.haystack.mit.edu/arrays/MWA/LFD} an interferometer to
be built in Western Australia, which will study polarised
emission over wide fields in the frequency range 80--300~MHz;
\item The extended New Technology Demonstrator
(xNTD)\footnote{http://www.atnf.csiro.au/projects/ska/xntd.html}
and the Karoo Array Telescope (KAT),\footnote{http://www.ska.ac.za/kat/} 
arrays to be
built in South Africa and Western Australia, respectively,
both of which will be
very wide-field (30--40~deg$^2$ at 1.4~GHz) survey instruments,
and which can study polarisation at a range of spatial scales
in the approximate frequency range 800--1700~MHz.
\end{itemize}
All these facilities are under construction, and should be operational
in the next 2--5 years. These telescopes will provide $\sim$100\,000
RMs  and measurements of diffuse polarisation all over the sky, allowing
many new studies of RM synthesis, polarisation silhouettes, and other
experiments discussed above.

\section{Conclusions}

``Cosmic Magnetism'' has been named as one of five key science projects
for the SKA. This telescope will open an entirely new regime for probing
magnetic fields at all redshifts, and will provide unique radio data that
can complement other information delivered on magnetic fields by Auger,
HESS, {\em Planck}\ and {\em GLAST}. The considerable new parameter space
opened up by the SKA implies that in addition to the experiments that
we can conceive today, the SKA will almost certainly discover entirely
new and unexpected magnetic phenomena.

In the years leading up to the construction and commissioning of the
SKA, we encourage theorists to include RM and polarisation predictions
in their calculations, and observers to start thinking about other
magnetic experiments that the SKA could carry out. In the meantime,
the magnetic field community can look forward to the SKA pathfinders
delivering a great deal of new data on the polarised sky.

\acknowledgements

B.M.G. acknowledges the support of the National Science Foundation
through grant AST-0307358.

\bibliographystyle{apj}
\bibliography{journals,modrefs,psrrefs,crossrefs}

\begin{thebibliography}{59}
\expandafter\ifx\csname natexlab\endcsname\relax\def\natexlab#1{#1}\fi

\bibitem[{{Bagchi} {et~al.}(2002){Bagchi}, {En{\ss}lin}, {Miniati}, {Stalin},
  {Singh}, {Raychaudhury}, \& {Humeshkar}}]{bem+02}
{Bagchi}, J., {En{\ss}lin}, T.A., {Miniati}, F., {Stalin}, C.S., {Singh}, M.,
  {Raychaudhury}, S.,  {Humeshkar}, N.B.: 2002, New Astr. 7, 249

\bibitem[{{Beck}(2005)}]{bec05}
{Beck}, R.: 2005, in: R.~Wielebinski, R.~Beck (eds.),
\emph{Cosmic Magnetic Fields}, Springer, Berlin, p.~41

\bibitem[{{Beck}(2006)}]{bec06}
{Beck}, R.: 2006, in: F.~Boulanger et al. (eds.),
\emph{Polarisation 2005}, in press


\bibitem[{{Beck} {et~al.}(1996){Beck}, {Brandenburg}, {Moss}, {Shukurov}, \&
  {Sokoloff}}]{bbm+96}
{Beck}, R., {Brandenburg}, A., {Moss}, D., {Shukurov}, A., {Sokoloff}, D.:
  1996, ARA\&A 34, 155

\bibitem[{Beck \& Gaensler(2004)}]{bg04}
Beck, R., Gaensler, B.M.: 2004, New Astron. Rev. 48, 1289

\bibitem[{Becker {et~al.}(1995)Becker, White, \& Helfand}]{bwh95}
Becker, R.H., White, R.L., Helfand, D.J.: 1995, ApJ 450, 559

\bibitem[{Blasi {et~al.}(1999)Blasi, Burles, \& Olinto}]{bbo99}
Blasi, P., Burles, S., Olinto, A.V.: 1999, ApJ 514, L79

\bibitem[{Brentjens \& de~Bruyn(2005)}]{bd05}
Brentjens, M.A., de~Bruyn, A.G.: 2005, A\&A 441, 1217

\bibitem[{{Brown} {et~al.}(2003){Brown}, {Taylor}, \& {Jackel}}]{btj03}
{Brown}, J.C., {Taylor}, A.R., {Jackel}, B.J.: 2003, ApJS 145, 213

\bibitem[{Br\"{u}ggen {et~al.}(2005)Br\"{u}ggen, Ruszkowski, Simionescu, Hoeft,
  \& Dalla~Vecchi}]{brs+05}
Br\"{u}ggen, M., Ruszkowski, M., Simionescu, A., Hoeft, M., Dalla~Vecchi, C.:
  2005, ApJ 631, L21

\bibitem[{Carilli {et~al.}(2004{\natexlab{a}})Carilli, Furlanetto, Briggs,
  Jarvis, Rawlings, \& Falcke}]{cfb+04}
Carilli, C.L., Furlanetto, S., Briggs, F., Jarvis, M., Rawlings, S.,
  Falcke, H.: 2004{\natexlab{a}}, New Astron. Rev. 48, 1029

\bibitem[{Carilli \& Taylor(2002)}]{ct02}
Carilli, C.L., Taylor, G.B.: 2002, Ann. Rev. Astr. Ap. 40, 319

\bibitem[{Carilli {et~al.}(2004{\natexlab{b}})Carilli, Walter, Bertoldi,
  Menten, Fan, Lewis, Strauss, Cox, Beelen, Omont, \& Mohan}]{cwb+04}
Carilli, C.L., Walter, F., Bertoldi, F., et al.: 
  2004{\natexlab{b}}, AJ 128, 997

\bibitem[{Cordes {et~al.}(2004)Cordes, Kramer, Lazio, Stappers, Backer, \&
  Johnston}]{ckl+04}
Cordes, J.M., Kramer, M., Lazio, T.J.W., Stappers, B.W., Backer, D.C.,
  Johnston, S.: 2004, New Astr. Rev. 48, 1413

\bibitem[{Crutcher {et~al.}(2003)Crutcher, Heiles, \& Troland}]{cht03}
Crutcher, R., Heiles, C., Troland, T.: 2003, in: E.~Falgarone, T.~Passot
(eds.), \emph{Turbulence and Magnetic
  Fields in Astrophysics}, Springer, Berlin, p. 155

\bibitem[{de~Bruyn \& Brentjens(2005)}]{db05}
de~Bruyn, A.G., Brentjens, M.A.: 2005, A\&A 441, 931

\bibitem[{Feretti {et~al.}(2004)Feretti, Burigana, \& En{\ss}lin}]{fba04}
Feretti, L., Burigana, T., En{\ss}lin, T.A.: 2004, New Astron. Rev. 48,
  1137

\bibitem[{{Feretti} \& {Johnston-Hollitt}(2004)}]{fj04}
{Feretti}, L., {Johnston-Hollitt}, M.: 2004, New Astron. Rev. 48, 1145

\bibitem[{{Fomalont} {et~al.}(1989){Fomalont}, {Ebneter}, {van Breugel}, \&
  {Ekers}}]{feve89}
{Fomalont}, E.B., {Ebneter}, K.A., {van Breugel}, W.J.M.,  {Ekers}, R.D.:
  1989, ApJ 346, L17

\bibitem[{Furlanetto \& Loeb(2001)}]{fl01}
Furlanetto, S.R., Loeb, A.: 2001, ApJ 556, 619

\bibitem[{Gaensler {et~al.}(2004)Gaensler, Beck, \& Feretti}]{gbf04}
Gaensler, B.M., Beck, R., Feretti, L.: 2004, New Astron. Rev. 48, 1003

\bibitem[{{Gaensler} {et~al.}(2001){Gaensler}, {Dickey}, {McClure-Griffiths},
  {Green}, {Wieringa}, \& {Haynes}}]{gdm+01}
{Gaensler}, B.M., {Dickey}, J.M., {McClure-Griffiths}, N.M., {Green}, A.J.,
  {Wieringa}, M.H., {Haynes}, R.F.: 2001, ApJ 549, 959

\bibitem[{Gaensler {et~al.}(2005)Gaensler, Haverkorn, Staveley-Smith, Dickey,
  McClure-Griffiths, Dickel, \& Wolleben}]{ghs+05}
Gaensler, B.M., Haverkorn, M., Staveley-Smith, L., Dickey, J.M.,
  McClure-Griffiths, N.M., Dickel, J.R., Wolleben, M.: 2005, Science 307,
  1610

\bibitem[{Gaustad {et~al.}(2001)Gaustad, McCullough, Rosing, \&
  Van~Buren}]{gmrv01}
Gaustad, J.E., McCullough, P.R., Rosing, W., Van~Buren, D.: 2001, PASP
  113, 1326

\bibitem[{Govoni {et~al.}(2005)Govoni, Murgia, Feretti, Giovannini, Dallacasa,
  \& Taylor}]{gmf+05}
Govoni, F., Murgia, M., Feretti, L., Giovannini, G., Dallacasa, D., Taylor,
  G.B.: 2005, A\&A 430, L5

\bibitem[{{Govoni} {et~al.}(2001){Govoni}, {Taylor}, {Dallacasa}, {Feretti}, \&
  {Giovannini}}]{gtd+01}
{Govoni}, F., {Taylor}, G.B., {Dallacasa}, D., {Feretti}, L., {Giovannini},
  G.: 2001, A\&A 379, 807

\bibitem[{{Han} {et~al.}(1998){Han}, {Beck}, \& {Berkhuijsen}}]{hbb98}
{Han}, J.L., {Beck}, R., {Berkhuijsen}, E.M.: 1998, A\&A 335, 1117

\bibitem[{Han {et~al.}(2006)Han, Manchester, Lyne, Qiao, \& van
  Straten}]{hml+06}
Han, J.L., Manchester, R.N., Lyne, A.G., Qiao, G.J., van Straten, W.:
  2006, ApJ in press (astro-ph/0601357)

\bibitem[{{Hopkins} {et~al.}(2003){Hopkins}, {Afonso}, {Chan}, {Cram},
  {Georgakakis}, \& {Mobasher}}]{hac+03}
{Hopkins}, A.M., {Afonso}, J., {Chan}, B., {Cram}, L.E., {Georgakakis}, A.,
 {Mobasher}, B.: 2003, AJ 125, 465

\bibitem[{Jackson(2004)}]{jac04}
Jackson, C.A.: 2004, New Astron. Rev. 48, 1187

\bibitem[{Jedamzik {et~al.}(2000)Jedamzik, Katalini\'{c}, \& Olinto}]{jko00}
Jedamzik, K., Katalini\'{c}, V., Olinto, A.V.: 2000, Phys. Rev. Lett. 85,
  700

\bibitem[{{Johnson} {et~al.}(1995){Johnson}, {Leahy}, \& {Garrington}}]{jlg95}
{Johnson}, R.A., {Leahy}, J.P., {Garrington}, S.T.: 1995, MNRAS 273, 877

\bibitem[{Jones(2004)}]{jon04}
Jones, D.L.: 2004, SKA Science Requirements, SKA Memo Series, No. 45

\bibitem[{Kawai {et~al.}(2005)Kawai, Yamada, Kosugi, Hattori, \& Aoki}]{kyk+05}
Kawai, N., Yamada, T., Kosugi, G., Hattori, T., Aoki, K.: 2005, GCN Circular
  3937

\bibitem[{Keshet {et~al.}(2004{\natexlab{a}})Keshet, Waxman, \& Loeb}]{kwl04a}
Keshet, U., Waxman, E., Loeb, A.: 2004{\natexlab{a}}, ApJ 617, 281

\bibitem[{Keshet {et~al.}(2004{\natexlab{b}})Keshet, Waxman, \& Loeb}]{kwl04b}
Keshet, U., Waxman, E., Loeb, A.: 2004{\natexlab{b}}, New Astr. 48, 1119

\bibitem[{Kolatt(1998)}]{kol98}
Kolatt, T.: 1998, ApJ 495, 564

\bibitem[{Kosowsky {et~al.}(2005)Kosowsky, Kashniashvili, Lavrelashvili, \&
  Ratra}]{kklr05}
Kosowsky, A., Kashniashvili, T., Lavrelashvili, G.,  Ratra, B.: 2005, Phys.
  Rev. D 71, 043006

\bibitem[{Kramer {et~al.}(2004)Kramer, Backer, Lazio, Stappers, \&
  Johnston}]{kbc+04}
Kramer, M., Backer, D.C., Lazio, T. J.W., Stappers, B.W., Johnston, S.:
  2004, New Astron. Rev. 48, 993

\bibitem[{{Kronberg}(1994)}]{kro94}
{Kronberg}, P.P.: 1994, Rep. Prog. Phys. 57, 325

\bibitem[{{Kronberg} {et~al.}(1992){Kronberg}, {Perry}, \& {Zukowski}}]{kpz92}
{Kronberg}, P.P., {Perry}, J.J., {Zukowski}, E.L.H.: 1992, ApJ 387, 528

\bibitem[{{Kulsrud}(1999)}]{kul99}
{Kulsrud}, R.M.: 1999, ARA\&A 37, 37

\bibitem[{Lazio {et~al.}(1990)Lazio, Spangler, \& Cordes}]{lsc90}
Lazio, T.J., Spangler, S.R., Cordes, J.M.: 1990, ApJ 363, 515

\bibitem[{Lazio {et~al.}(2004)Lazio, Tarter, \& Wilner}]{ltw04}
Lazio, T.J.W., Tarter, J.C.,  Wilner, D.J.: 2004, New Astron. Rev. 48,
  985

\bibitem[{Minter \& Spangler(1996)}]{ms96b}
Minter, A.H., Spangler, S.R.: 1996, ApJ 458, 194

\bibitem[{Oren \& Wolfe(1995)}]{ow95}
Oren, A.L., Wolfe, A.M.: 1995, ApJ 445, 624

\bibitem[{Perry {et~al.}(1993)Perry, Watson, \& Kronberg}]{pwk93}
Perry, J.J., Watson, A.M., Kronberg, P.P.: 1993, ApJ 406, 407

\bibitem[{Rawlings {et~al.}(2004)Rawlings, Abdalla, Bridle, Blake, Baugh,
  Greenhill, \& van~der Hulst}]{rab+04}
Rawlings, S., Abdalla, F.B., Bridle, S.L., Blake, C.A., Baugh, C.M.,
  Greenhill, L.J., van~der Hulst, J.M.: 2004, New Astron. Rev. 48, 1013

\bibitem[{{Ruzmaikin} {et~al.}(1988){Ruzmaikin}, {Sokolov}, \&
  {Shukurov}}]{rss88}
{Ruzmaikin}, A.A., {Sokolov}, D.D., {Shukurov}, A.M.: 1988, \emph{Magnetic
  Fields of Galaxies}, Kluwer, Dordrecht

\bibitem[{Schilizzi(2006)}]{sch06}
Schilizzi, R.T.: 2006, Reference design for the SKA, SKA Memo Series, No. 69

\bibitem[{Schulman \& Fomalont(1992)}]{sf92}
Schulman, E., Fomalont, E.B.: 1992, AJ 103, 1138

\bibitem[{Sethi \& Subramanian(2005)}]{ss05}
Sethi, S.K., Subramanian, K.: 2005, MNRAS 356, 778

\bibitem[{Shukurov(2005)}]{shu05}
Shukurov, A. 2005, in: R.~Wielebinski, R.~Beck (eds.),
\emph{Cosmic Magnetic Fields}, Springer, Berlin, p. 113

\bibitem[{Sokoloff {et~al.}(1998)Sokoloff, Bykov, Shukurov, Berkhuijsen, Beck,
  \& Poezd}]{sbs+98}
Sokoloff, D.D., Bykov, A.A., Shukurov, A., Berkhuijsen, E.M., Beck, R., 
  Poezd, A.D.: 1998, MNRAS 299, 189

\bibitem[{{Stepanov} {et~al.}(2002){Stepanov}, {Frick}, {Shukurov}, \&
  {Sokoloff}}]{sfss02}
{Stepanov}, R., {Frick}, P., {Shukurov}, A., {Sokoloff}, D.: 2002, A\&A 391,
  361

\bibitem[{Vogt \& En{\ss}lin(2003)}]{ve03}
Vogt, C., En{\ss}lin, T.A.: 2003, A\&A 412, 373

\bibitem[{Wasserman(1978)}]{was78}
Wasserman, I.: 1978, ApJ 224, 337

\bibitem[{{Welter} {et~al.}(1984){Welter}, {Perry}, \& {Kronberg}}]{wpk84}
{Welter}, G.L., {Perry}, J.J., {Kronberg}, P.P.: 1984, ApJ 279, 19

\bibitem[{Wilkinson {et~al.}(2004)Wilkinson, Kellermann, Ekers, Cordes, \&
  Lazio}]{wke+04}
Wilkinson, P.N., Kellermann, K.I., Ekers, R.D., Cordes, J.M., Lazio, T.J.W.: 
2004, New Astron. Rev. 48, 1551

\bibitem[{Xu {et~al.}(2006)Xu, Kronberg, Habib, \& Dufton}]{xkhd06}
Xu, Y., Kronberg, P.P., Habib, S., Dufton, Q.W.: 2006, ApJ 637, 19

\end{thebibliography}


\end{document}